# 3-Uniform states and orthogonal arrays

**Xin-wei Zha**[1,*], **Irfan Ahmed**[2,3], **Yanpeng Zhang**[2,†]

[1]School of Science, Xi'an University of Posts and Telecommunications, Xi'an, 710121, China

[2]Key Laboratory for Physical Electronics and Devices of the Ministry of Education & Shaanxi Key Lab of Information Photonic Technique, Xi'an Jiaotong University, Xi'an 710049, China

[3]Department of Electrical Engineering, Sukkur IBA, Sukkur 65200, Sindh, Pakistan

Corresponding author: *zhxw@xupt.edu.cn; †ypzhang@mail.xjtu.edu.cn

**Abstract:** In a recent paper (Phys. Rev. A **90**, 022316 (2014) ), Goyeneche *et al.* established a link between the combinatorial notion of orthogonal arrays and *k*-uniform states and present open issue. '(B) Find for what *N* there are 3-uniform states of *N*-qubits'. In this paper, we demonstrate the existence of 3-uniform states of  *N*-qubits for N=11,..,15".

**1. Introduction**: Entanglement is considered as the key and central resources for quantum information and computation [1, 2], numerous theoretical and experimental works have been done in this field [3–6]. Since last decades, a lot of efforts have been made to qualitatively quantify the amount of entanglement of various multipartite states. In particular, search for maximally entangled states has been focused with great attention [7–12]. In case of 2-qubits, it is known that Bell states are maximally entangled with respect to any measures of entanglement. For higher numbers of qubits, the problem is no longer simple. As an important feature in quantum many-body systems, multi-particle entanglement has been intensively investigated since it is found significantly different from the trivial extension of bipartite entanglement [13–17]. Recently, Goyeneche and Życzkowski established a link between the combinatorial notion of orthogonal arrays and k-uniform states and constructed 2-uniform states for an arbitrary number of N > 5-qubits [18]. They also list open issue such as (B) Find for what *N* there are 3-uniform states of *N*-qubits?

In this article, we answer the open issue as mentioned in the abstract in subsequent sections. In second section, we propose a criterion for the *k*-uniform state via local unitary transformation invariant. In third, we give the 3-uniform states for 11, 12,13,14,15-qubits. In last section we conclude the article.

## 2. Criterion of *k*-uniform state

An *n*-qubit pure state $|\psi\rangle$ is a *k*-uniform states if all its reductions to *k* qubits are maximally mixed [18,19]. On the other hand, we may give

$$\pi_i = Tr_i \rho_i^2 = \frac{1}{2} + \frac{1}{2} F_i,$$
$$i = 1, 2, \cdots, n$$

(1)

$$\pi_{ij} = Tr_{ij} \rho_{ij}^2 = \frac{1}{4} + \frac{1}{4}\left(F_i + F_j + F_{ij}\right),$$
$$ij = 12, 13, \cdots n-1, n$$

(2)

$$\pi_{ijk} = Tr_{ijk} \rho_{ijk}^2 = \frac{1}{8} + \frac{1}{8}\left(F_i + F_j + F_k + F_{ij} + F_{ik} + F_{jk} + F_{ijk}\right),$$
$$ijk = 123, 124, \cdots, n-2, n-1, n$$

(3)

Where

$$F_i = \langle\psi|\hat{\sigma}_{ix}|\psi\rangle^2 + \langle\psi|\hat{\sigma}_{iy}|\psi\rangle^2 + \langle\psi|\hat{\sigma}_{iz}|\psi\rangle^2$$

$$F_{ij} = \langle\psi|\hat{\sigma}_{ix}\hat{\sigma}_{jx}|\psi\rangle^2 + \langle\psi|\hat{\sigma}_{ix}\hat{\sigma}_{jy}|\psi\rangle^2 + \langle\psi|\hat{\sigma}_{ix}\hat{\sigma}_{jz}|\psi\rangle^2$$
$$+ \langle\psi|\hat{\sigma}_{iy}\hat{\sigma}_{jx}|\psi\rangle^2 + \langle\psi|\hat{\sigma}_{iy}\hat{\sigma}_{jy}|\psi\rangle^2 + \langle\psi|\hat{\sigma}_{iy}\hat{\sigma}_{jz}|\psi\rangle^2$$
$$+ \langle\psi|\hat{\sigma}_{iz}\hat{\sigma}_{jx}|\psi\rangle^2 + \langle\psi|\hat{\sigma}_{iz}\hat{\sigma}_{jy}|\psi\rangle^2 + \langle\psi|\hat{\sigma}_{iz}\hat{\sigma}_{jz}|\psi\rangle^2$$

$$F_{ijk} = \langle\psi|\hat{\sigma}_{ix}\hat{\sigma}_{jx}\hat{\sigma}_{kx}|\psi\rangle^2 + \langle\psi|\hat{\sigma}_{ix}\hat{\sigma}_{jx}\hat{\sigma}_{ky}|\psi\rangle^2 + \langle\psi|\hat{\sigma}_{ix}\hat{\sigma}_{jx}\hat{\sigma}_{kz}|\psi\rangle^2$$
$$+ \langle\psi|\hat{\sigma}_{ix}\hat{\sigma}_{jy}\hat{\sigma}_{kx}|\psi\rangle^2 + \langle\psi|\hat{\sigma}_{ix}\hat{\sigma}_{jy}\hat{\sigma}_{ky}|\psi\rangle^2 + \langle\psi|\hat{\sigma}_{ix}\hat{\sigma}_{jy}\hat{\sigma}_{kz}|\psi\rangle^2$$
$$+ \cdots$$
$$+ \langle\psi|\hat{\sigma}_{iz}\hat{\sigma}_{jz}\hat{\sigma}_{kx}|\psi\rangle^2 + \langle\psi|\hat{\sigma}_{iz}\hat{\sigma}_{jz}\hat{\sigma}_{ky}|\psi\rangle^2 + \langle\psi|\hat{\sigma}_{iz}\hat{\sigma}_{jz}\hat{\sigma}_{kz}|\psi\rangle^2$$

It is obvious that such invariants satisfy $F_i \geq 0, F_{ij} \geq 0, F_{ijk} \geq 0$. From Eq. (1), we know, if $F_i = 0$, we have $\pi_i = Tr_i \rho_i^2 = \frac{1}{2}, i = 1, 2, \cdots, n$. Namely, every reduced one-qubit state is completely mixed. Therefore, this state is a 1-uniform state.

From Eq. (2), we know, if $F_i = 0$, $F_j = 0$ and $F_{ij} = 0$, then we have $\pi_{ij} = Tr_{ij}\rho_{ij}^2 = \frac{1}{4}$. This state is a 2-uniform state.

Similarly, from Eq. (3), we know, if $F_i = 0$, $F_j = 0$, $F_k = 0$ $F_{ij} = 0$, $F_{ik} = 0$, $F_{jk} = 0$ and $F_{ijk} = 0$, then we have $\pi_{ijk} = Tr_{ijk}\rho_{ijk}^2 = \frac{1}{8}$. This state is a 3-uniform state. Obviously, a $k$-uniform state is also a $m$-uniform for $m < k$.

## 3. Issue B: 3-uniform states of $N$-qubits generated from an OA

Goyeneche and Życzkowski used known Hadamard matrices to exemplify the construction of 2-uniform states for $N$ = 8, 9,10,11,12,13,14,15. In this section we solve the problem of constructing a kind of 3-uniform states for number of 11-qubits. Using Eqs (1-3), we can find 3-uniform states for 11-qubits.

$$\begin{aligned}|\psi_{11}^2\rangle = \frac{1}{4\sqrt{2}}[&(|0000\rangle + |1111\rangle)_{3467}(|0000000\rangle + |1111111\rangle)_{12589,10,11} \\
&+ (|0001\rangle + |1110\rangle)_{3467}(|0111010\rangle + |1000101\rangle)_{12589,10,11} \\
&+ (|0010\rangle + |1101\rangle)_{3467}(|0110101\rangle + |1001010\rangle)_{12589,10,11} \\
&+ (|0011\rangle + |1100\rangle)_{3467}(|0001111\rangle + |1110000\rangle)_{12589,10,11} \\
&+ (|0100\rangle + |1011\rangle)_{3467}(|0101100\rangle + |1010011\rangle)_{12589,10,11} \\
&+ (|0101\rangle + |1010\rangle)_{3467}(|0010110\rangle + |1101001\rangle)_{12589,10,11} \\
&+ (|0110\rangle + |1001\rangle)_{3467}(|0011001\rangle + |1100110\rangle)_{12589,10,11} \\
&+ (|0111\rangle + |1000\rangle)_{3467}(|0100011\rangle + |1011100\rangle)_{12589,10,11}]\end{aligned} \quad (4)$$

We can show that $F_i = 0$, i=1,2,$\cdots$,11, $F_{ij} = 0, ij = 12,13,\cdots,(10,11)$,

$F_{ijk} = 0, ijk = 123,124,\cdots,(9,10,11)$. And $\pi_{ijk} = Tr_{ijk}\rho_{ijk}^2 = \frac{1}{8}, ijk = 123,124,\cdots,(9,10,11)$.

Therefore, this state is a 3-uniform state for 11-qubits.

Similarly, we can also obtain the 3-uniform state of 12,13,14,15-qubits

$$|\psi_{12}\rangle = \frac{1}{4\sqrt{2}}[(|0000\rangle+|1111\rangle)_{3467}(|00000000\rangle+|11111111\rangle)_{12589,10,11,12}$$
$$+(|0001\rangle+|1110\rangle)_{3467}(|01110100\rangle+|10001011\rangle)_{12589,10,11,12}$$
$$+(|0010\rangle+|1101\rangle)_{3467}(|01101010\rangle+|10010101\rangle)_{12589,10,11,12}$$
$$+(|0011\rangle+|1100\rangle)_{3467}(|00011110\rangle+|11100001\rangle)_{12589,10,11,12}$$
$$+(|0100\rangle+|1011\rangle)_{3467}(|01011001\rangle+|10100110\rangle)_{12589,10,11,12}$$
$$+(|0101\rangle+|1010\rangle)_{3467}(|00101101\rangle+|11010010\rangle)_{12589,10,11,12}$$
$$+(|0110\rangle+|1001\rangle)_{3467}(|00110011\rangle+|11001100\rangle)_{12589,10,11,12}$$
$$+(|0111\rangle+|1000\rangle)_{3467}(|01000111\rangle+|10111000\rangle)_{12589,10,11,12}]$$

(5)

$$|\psi_{13}\rangle = \frac{1}{4\sqrt{2}}[|0000\rangle_{128,11}(|000000000\rangle+|011011110\rangle+|101101111\rangle+|110110001\rangle)_{34567910,12,13}$$
$$+|0011\rangle_{128,11}(|000111101\rangle+|011100011\rangle+|101010010\rangle+|110001100\rangle)_{34567910,12,13}$$
$$+|0101\rangle_{128,11}(|001101000\rangle+|010110110\rangle+|100000111\rangle+|111011001\rangle)_{34567910,12,13}$$
$$+|0110\rangle_{128,11}(|001010101\rangle+|010001011\rangle+|100111010\rangle+|111100100\rangle)_{34567910,12,13}$$
$$+|1001\rangle_{128,11}(|000011011\rangle+|011000101\rangle+|101110100\rangle+|110101010\rangle)_{34567910,12,13}$$
$$+|1010\rangle_{128,11}(|000100110\rangle+|011111000\rangle+|101001001\rangle+|110010111\rangle)_{34567910,12,13}$$
$$+|1100\rangle_{128,11}(|001110011\rangle+|010101101\rangle+|100011100\rangle+|111000010\rangle)_{34567910,12,13}$$
$$+|1111\rangle_{128,11}(|001001110\rangle+|010010000\rangle+|100100001\rangle+|111111111\rangle)_{34567910,12,13}]$$

(6)

$$|\psi_{14}\rangle = \frac{1}{4\sqrt{2}}[|0000\rangle_{128,11}(|0000000000\rangle+|0110111101\rangle+|1011011110\rangle+|1101100011\rangle)_{34567910,12,13,14}$$
$$+|0011\rangle_{128,11}(|0001111011\rangle+|0111000110\rangle+|1010100101\rangle+|1100011000\rangle)_{34567910,12,13,14}$$
$$+|0101\rangle_{128,11}(|0011010001\rangle+|0101101100\rangle+|1000001111\rangle+|1110110010\rangle)_{34567910,12,13,14}$$
$$+|0110\rangle_{128,11}(|0010101010\rangle+|0100010111\rangle+|1001110100\rangle+|1111001001\rangle)_{34567910,12,13,14}$$
$$+|1001\rangle_{128,11}(|0000110110\rangle+|0110001011\rangle+|1011101000\rangle+|1101010101\rangle)_{34567910,12,13,14}$$
$$+|1010\rangle_{128,11}(|0001001101\rangle+|0111110000\rangle+|1010010011\rangle+|1100101110\rangle)_{34567910,12,13,14}$$
$$+|1100\rangle_{128,11}(|0011100111\rangle+|0101011010\rangle+|1000111001\rangle+|1110000100\rangle)_{34567910,12,13,14}$$
$$+|1111\rangle_{128,11}(|0010011100\rangle+|0100100001\rangle+|1001000010\rangle+|1111111111\rangle)_{34567910,12,13,14}]$$

(7)

$$|\psi_{15}\rangle = \frac{1}{4\sqrt{2}}[|0000\rangle_{128,11}(|00000000000\rangle+|01101111010\rangle+|10110111101\rangle+|11011000111\rangle)_{34567910,12,13,14,15}$$
$$+|0011\rangle_{128,11}(|00011110110\rangle+|01110001100\rangle+|10101001011\rangle+|11000110001\rangle)_{34567910,12,13,14,15}$$
$$+|0101\rangle_{128,11}(|00110100011\rangle+|01011011001\rangle+|10000011110\rangle+|11101100100\rangle)_{34567910,12,13,14,15}$$
$$+|0110\rangle_{128,11}(|00101010101\rangle+|01000101111\rangle+|10011101000\rangle+|11110010010\rangle)_{34567910,12,13,14,15}$$
$$+|1001\rangle_{128,11}(|00001101101\rangle+|01100010111\rangle+|10111010000\rangle+|11010101010\rangle)_{34567910,12,13,14,15}$$
$$+|1010\rangle_{128,11}(|00010011011\rangle+|01111100001\rangle+|10100100110\rangle+|11001011100\rangle)_{34567910,12,13,14,15}$$
$$+|1100\rangle_{128,11}(|00111001110\rangle+|01010110100\rangle+|10001110011\rangle+|11100001001\rangle)_{34567910,12,13,14,15}$$
$$+|1111\rangle_{128,11}(|00100111000\rangle+|01001000010\rangle+|10010000101\rangle+|11111111111\rangle)_{34567910,12,13,14,15}]$$

(8)

As reference [18], sometimes it is possible to introduce *minus signs mathematically* in some terms of the state such that it becomes a *k*-uniform state. But practically, Quantum state tomography (QST) through quantum logic gates can be achieved through quantum state of logic operation particularly Hadamard phase shift (Paulis Z-gate) operation of qubits. Using this method, we can get a 3-uniform state of 8-qubits [12]. Then, we have

$$|\psi_M\rangle_{12345678} =$$
$$\frac{1}{8}\{[(|000\rangle+|111\rangle)|0\rangle+(|010\rangle+|101\rangle)|1\rangle][(|000\rangle-|111\rangle)|0\rangle+(|000\rangle-|111\rangle)|1\rangle]$$
$$+[(|001\rangle+|110\rangle)|0\rangle+(|000\rangle-|111\rangle)|1\rangle][(|000\rangle+|111\rangle)|0\rangle+(|100\rangle+|011\rangle)|1\rangle] \quad (9)$$
$$+[(|010\rangle+|101\rangle)|0\rangle+(|100\rangle+|011\rangle)|1\rangle][(|000\rangle+|111\rangle)|0\rangle-(|010\rangle+|101\rangle)|1\rangle]$$
$$+[(|000\rangle-|111\rangle)|0\rangle+(|010\rangle-|101\rangle)|1\rangle][(|000\rangle-|111\rangle)|0\rangle+(|100\rangle-|101\rangle)|1\rangle]\}$$

We have $\pi_{ijk} = Tr_{ijk}\rho_{ijk}^2 = \frac{1}{8}, ijk = 123,124,\cdots,9,10,11$

This state is the 3-uniform state of 8-qubits.

We can also obtain

$$\pi_{1236} = \frac{1}{4}, \pi_{1245} = \frac{1}{4}, \pi_{1278} = \frac{1}{4};$$

$$\pi_{1348} = \frac{1}{8}, \pi_{1357} = \frac{1}{8}, \pi_{1467} = \frac{1}{8}, \pi_{1568} = \frac{1}{8}$$

$$\pi_{ijkl} = \frac{1}{16}, ijkl = 1234,1235,1237,\cdots,1678$$

(10)

The advantages of 3-uniform over 2-uniform state is that we may increase the order of density coding information rate from $2^2$ to $2^3$. This may increase the quantum error-correcting codes (QECCs) as K-uniform states of N-qubits exist for only $k \in \mathbf{N}$ if N is sufficiently large [18], for example [20] predicts 3 uniform state for $N \geq 14$ qubits. The two and three uniform state of entangled beam can be viewed randomly and their order of density coding information rate can also be increased with this kind of uniformity as shown below.

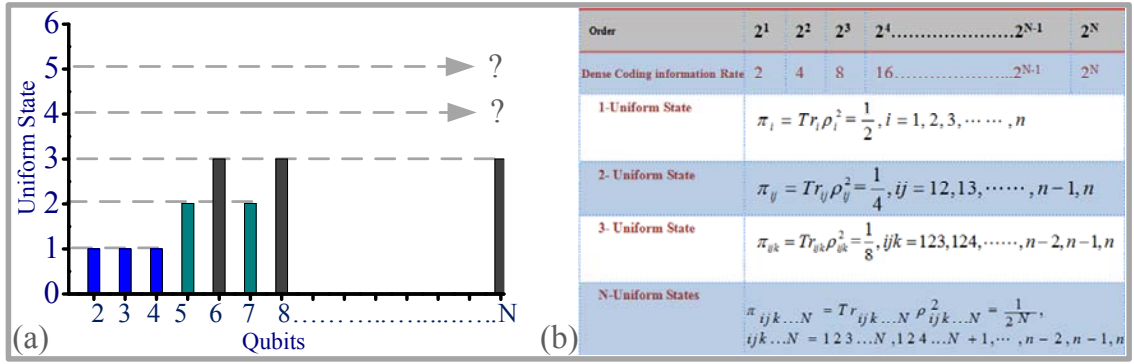

Figure 1(a) shows the uniform state of qubits 2,3.4,5,6,7,8…………..N. (b) Shows the table that predicts order of increment in dense coding information rate with order of uniform state.

In Fig. 1(a), we have shown 1-uniform state for 2,3,4-qubits, 2-uniform state for 5,7-qubits [18], 3-uniform state for 6,8-qubits [12] and 3-uniform state for 11,12,13,14,15-qubits we have found above and can be generalized to N-qubits. While 4 and 5-unifrom states for these qubits are not defined yet. Figure 1(b) shows the table that predicts the order of increment in dense coding information rate with order of uniform state.

### 4. Conclusion

In conclusion, we found that a 3-uniform states for 11, 12,13,14,15 qubits. We may also find 1-uniform state for 2,3,4-qubits, 2-uniform states for 5,7-qubits, 3-uniform states for 6,8,11,12,13,14,15-qubits by introducing *minus signs* in some terms. Beside these, advantages of three uniform state over two could be increase in the order of density coding information rate from $2^2$ to $2^3$. We believe those result can play an important role in quantum communication and computing.

### Acknowledgments

This work is supported by the Shaanxi Natural Science Foundation under Contract No.

2009JM1007.